\documentclass[useAMS,usenatbib]{mn2e}
\usepackage{graphicx}
\def\n85{NGC~185}
  
\def\kms{\relax \ifmmode {\,\rm km\,s}^{-1}\else \,km\,s$^{-1}$\fi}

\def\mincir{\ \raise-2.truept\hbox{\rlap{\hbox{$\sim$}}\raise5.truept
    \hbox{$<$}\ }}
\def\magcir{\ \raise-2.truept\hbox{\rlap{\hbox{$\sim$}}\raise5.truept
    \hbox{$>$}\ }}

\def\arcmin{$'$}

\def\ha{H$\alpha$}

\title[The Ionization Mechanism of NGC~185 ]{The Ionization Mechanism of NGC~185: How to Fake a Seyfert Galaxy
\thanks{Based on observations obtained at the Gemini 
Observatory, which is operated by the Association of Universities for Research in Astronomy, 
Inc., under a cooperative agreement with the NSF on behalf of the Gemini partnership.}}

\author[Martins et al.]
{L. P. Martins$^{1}$\thanks{E-mail:
lucimara.martins@cruzeirodosul.edu.br}, G. Lanfranchi$^{1}$, D. R. Gon\c calves$^{2,3}$, L. Magrini$^{4}$, \newauthor A. M. Teodorescu$^{5}$ and C. Quireza$^{2}$
\\
  $^{1}$ NAT - Universidade Cruzeiro do Sul, Rua Galv\~ao Bueno 868, 01506-000 S\~ao Paulo, Brazil\\
  $^{2}$ UFRJ - Observat\'orio do Valongo, Ladeira Pedro Antonio 43, 20080-090 Rio de Janeiro, Brazil\\
  $^{3}$ Department of Physics and Astronomy, University College London, Gower Street, WC1E 6BT  London, UK\\
  $^{4}$ INAF - Osservatorio Astrofisico di Arcetri, Largo E. Fermi 5, I-50125 Firenze, Italy\\
  $^{5}$ Institute for Astronomy, University of Hawaii, 2680 Woodlawn Drive, HI 96822 Honolulu, USA\\
}

\begin{document}

\date{Accepted ?. Received ?; in original form ?}

\pagerange{\pageref{firstpage}--\pageref{lastpage}} \pubyear{2011}

\maketitle

\label{firstpage}

%%%%%%%%%%%%%%%%%%%%%%%%%%%%%%%%%%%%%%%%%%%%%%%%%%%%%%%%%%%%%%%%%%%%%%%%%
\begin{abstract}

 NGC 185 is a dwarf spheroidal satellite of the Andromeda galaxy. From mid-1990s
onward it was revealed that dwarf spheroidals often display a varied and in some cases
complex star formation history. In an optical survey of bright nearby galaxies, 
NGC~185 was classified as a Seyfert galaxy based on its emission line ratios. However,
although the emission lines in this object formally place it in the category of Seyferts
it is probable that this galaxy does not contain a genuine active nucleus. NGC~185 was
not detected in radio surveys either in 6 cm or 20 cm, or x-ray observations,
which means that the Seyfert-like 
line ratios may be produced by stellar processes. In this work we try to identify
the possible ionization mechanisms for this galaxy. We discussed the possibility of the
line emissions being produced by planetary nebulae (PNe), using deep spectroscopy observations 
obtained with GMOS-N (Gemini Multi-Object Spectrograph - North), at Gemini. Although the flux of the PNe are high enough to explain the integrated
spectrum, the line ratios are very far from the values for the Seyfert classification. We then proposed
that a mixture of supernova remnants and PNe could be the source of the ionization,
and we show that a composition of these two objects do mimic seyfert-like line ratios.
We used chemical evolution models to predict the supernova rates and to 
support the idea that these supernova
remnants should be present in the galaxy.

\end{abstract}

\begin{keywords}

Galaxies: abundances - evolution - Local Group - Individual (NGC~185); ISM: Planetary Nebulae - supernova remnants
\end{keywords}

%%%%%%%%%%%%%%%%%%%%%%%%%%%%%%%%%%%%%%%%%%%%%%%%%%%%%%%%%%%%%%%%%%%%%%%%%
\section[]{Introduction}

NGC~185 is a dwarf spheroidal (dSph) and together with NGC~205 and NGC~147, is one of the three brightest dwarf companions of M31. 
Unlike the other two however, NGC~185 has an important content 
of gas and dust (Marleau, Noriega-Crespo \& Misselt 2010). 
At first,
the dwarf ellipticals (dEs) and the dSphs have been regarded as simple, old stellar systems, with a stellar content
resembling that of Galactic globular clusters. However, from mid-1990s onward, with the 
advent of Hubble Space Telescope, it was revealed that those simple galaxies often
display a varied and in some cases complex star formation history (\citealt{mateo98}).
A dozen of luminous blue stars (\citealt{baade51}) and other Population I features 
(dust clouds, HI gas, supernova remnants)
were detected at the center of NGC~185. This has been an intriguing feature of this galaxy for several decades. 
This recent star formation is confined
to its central 150 pc $\times$ 90 pc where the youngest (100 Myr old) population
is found (\citealt{lea93}; Mart\'i nez-Delgado, Aparicio \& Gallart 1999). 

In an optical survey of bright nearby galaxies carried out by Ho et al. (1997, hereafter HO97), 
NGC 185 was classified as a Seyfert galaxy based on its emission line ratios. 
This is somewhat unexpected since active galactic nuclei (AGNs) are not commonly found in these small
host galaxies. If true, this would be the only dSph with an active nucleus known today.
The nucleus of NGC~185 is very ill defined, and the spectrum they obtained has low S/N.
The emission lines of the galaxy are very weak and although
their ratio formally place it in the category
of Seyferts it is probable 
that this galaxy does not contain a genuine active nucleus. NGC 185 was not detected in radio
surveys either in 6 cm or 20 cm (\citealt{houlv01}, \citealt{hbc80}), or
x-ray observations (\citealt{brandt97}), which
means that the Seyfert-like line ratios may be produced
by stellar processes.

In this work we try to unveil the possible ionization mechanisms of NGC~185. In \S 2
we discuss its Seyfert classification using the integrated spectrum from HO97.
In \S 3 we study the possibility that planetary nebulae are responsible for the galaxy's emission
lines. To do this we used planetary nebulae's observations obtained by our group with Gemini.
In \S 4 we propose another ionization mechanism for this galaxy, based on diagnostic
diagrams. In \S 5 we support our theory using a chemical model of the galaxy and 
in \S 6 we present 
our conclusions.

%%%%%%%%%%%%%%%%%%%%%%%%%%%%%%%%%%%%%%%%%%%%%%%%%%%%%%%%%%%%%%%%%%%%%%%%%
\section[]{The Seyfert Classification}

HO97 classified NGC~185 as a Seyfert based on 
their spectrum of the central 2"x 4" of the galaxy. They 
acquired spectra covering the wavelength regions of 4230 - 5110~\AA ~and
6210 - 6860 \AA, with spectral resolutions of about 4 and
2.5~\AA~respectively. 

The optical spectral region contains several emission
lines whose intensity ratios can be used to discriminate different sources of ionization. 
\citet{bpt81} suggested a series of two-dimensional, line-intensity
ratio diagrams that have since become widely-used diagnostic tools to classify emission-line objects
(the BPT diagrams). 
In these diagrams, nebulae photoionized by hot, young stars (HII
regions) can be distinguished from those photoionized by a
harder radiation field, such as that of an AGN.
The separation between the two main ionization
sources (young stars vs. AGNs) and between the two AGN
excitation classes (Seyferts vs. Low Ionization Narrow Line Regions - LINERs) does not
have sharp, rigorously defined boundaries. In practice,
however, one must impose somewhat arbitrary, albeit
empirically motivated criteria to establish an internally
consistent system of classification. HO97
used the diagnostic diagrams recommended by \cite{vo87}, 
which employ optical line-intensity ratios
that are relatively insensitive to reddening  and were
contained in the spectral coverage of
their survey. The classification criteria adopted by them 
is presented in Table 1.

\begin{table*}
\centering
%\begin{minipage}{80mm}
\caption{Criteria for Spectral Classification }
\begin{tabular}{@{}ccccc@{}}
\hline
Class   & [OIII]$\lambda$5007/H$\beta$ & [OI]$\lambda$6300/H$\alpha$ &[NII]$\lambda$6584/H$\alpha$  & [SII]$\lambda\lambda$6716,6731/H$\alpha$    \\
\hline
HII nuclei       &  Any      &  $<$ 0.08            & $<$ 0.06    &  $<$ 0.04   \\ 
Seyfert nuclei   &  $\geq$ 3 & $\geq$ 0.08          & $\geq$ 0.06 &  $\geq$ 0.04 \\
LINERs           &  $<$ 3    & $\geq$ 0.17          & $\geq$ 0.06 &   $\geq$ 0.04\\
Transition nuclei&  $<$ 3    & $\geq$ 0.08, $<$ 0.17& $\geq$ 0.06 &   $\geq$ 0.04\\

\hline
\end{tabular}
%\end{minipage}
\label{criteria}
\end{table*}

The optical spectra of many emission-line nuclei are contaminated heavily,
and often dominated, by the stellar component. As late-type
giants dominate the integrated light of galaxy bulges, stellar
absorption lines affect the strengths of most emission lines
of interest. The magnitude of the effect obviously depends
on the equivalent widths of the emission lines, but it is
generally large in the centers of most ``normal" galaxies,
where the emission lines are quite weak. Consequently,
quantitative studies of most emission-line nuclei depend
sensitively on the accuracy with which the stellar absorption lines can be removed.

To correct for this effect, HO97 adopted a modified version of the technique of
template subtraction: it subtracts from the
spectrum of interest a suitably scaled template spectrum
that best represents the continuum and absorption-line strength of the stellar component.
The resulting end
product should then be a continuum-subtracted, pure
emission-line spectrum. The template models adopted by HO97 were constructed
from a library of galaxy absorption-line spectra observed in their survey. For NGC~185
the template was constructed using one only galaxy, NGC 205, which
they assume to have very similar stellar population, but 
has no emission lines.

\subsection{Stellar Population Synthesis}

Since the starlight subtraction is so crucially important
to obtain correct emission-line fluxes, we decided to test HO97's template subtraction 
by applying a stellar population synthesis method to their NGC 185 spectra. 
For this we
used the code {\sc starlight} (\citealt{cea04}, \citealt{cea05}, \citealt{mateus06}, 
\citealt{asari07}, \citealt{cea09}). {\sc starlight} mixes computational techniques
originally developed for semi empirical population synthesis with 
ingredients from evolutionary synthesis models.
Basically the code fits an observed spectrum with a combination,
in different proportions, of a number of simple stellar populations (SSPs). 
We used a base of 45 SSPs in this work. These SSPs are from 
\cite{bc03}, and cover  15 ages
from 1 Myr to 13 Gyr, and three metallicities: z=0.004, 0.020 (solar) and 0.050. 
It is important to keep in mind that the synthesis is based on the whole 
spectrum fitting, and the limited wavelength coverage of this spectrum means results
have to be considered very carefully. Our aim here is not to describe
in details the stellar population and metallicity of the galaxy, but 
to discover how much the stellar population absorption can interfere 
in the emission line measurements.

\begin{figure} 
\includegraphics [width=85mm]{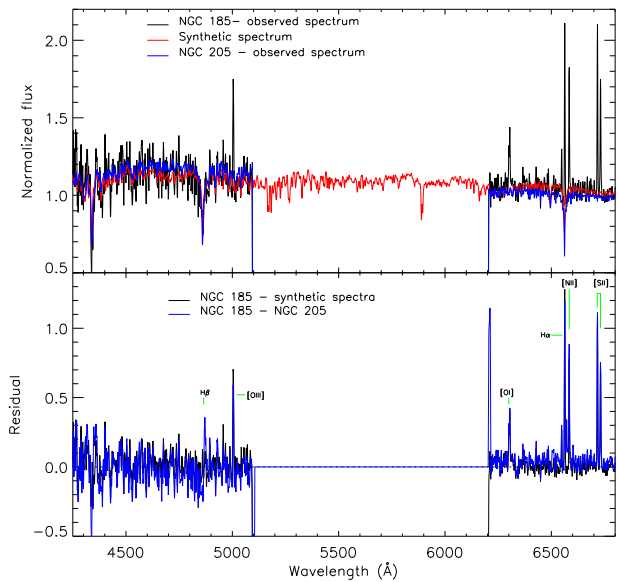}
\caption{Top panel: spectrum of NGC 185 from HO97 - black line. Superimposed it is shown 
the spectra of NGC 205, also from HO97, which was used by the authors as a 
stellar population template - blue line. In red is shown the result of the stellar population
synthesis of NGC 185 done with {\sc starlight}. Bottom panel: subtraction of the stellar population
of NGC~185 using the NGC~205 template (blue line) and using the synthetic spectra from
{\sc starlight}(black line) }
\label{starout}
\end{figure}

Figure~\ref{starout} shows the synthesis results. In the top panel is the
original HO97 spectrum (in black) and the 
synthetic spectra obtained by our stellar populations synthesis (in red). 
We also present the spectra of NGC~205 from HO97 for comparison
(in blue).  
As it turns out, the synthetic spectra is very similar to the spectra of NGC~205, which means
that HO97 line measurements were not affected by the template subtraction. The bottom 
panel shows the difference between the original and the synthetic
spectrum (black line) and between the original and NGC~205 spectrum (blue line).
It becomes clear that in both cases the H$\alpha$ line increases significantly after the 
subtraction, showing how important it is to correctly measure the absorption lines. H$\beta$
was not even detected in emission before the subtraction.

\begin{figure} 
\includegraphics [width=85mm]{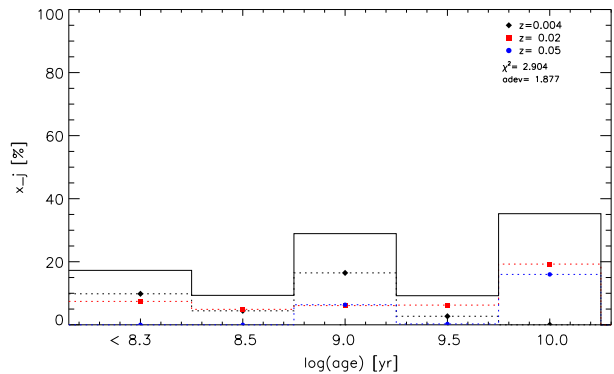}
\caption{Star formation history of NGC~185 obtained from the stellar population synthesis. 
The plot shows the population vector in flux fraction x$_j$ as a function of the 5 age bins.
The symbols represent
the fraction for each metallicity, and the black solid line represent the sum for
all three metallicities. }
\label{starage}
\end{figure}

From the stellar population synthesis we can obtain the
star formation history (SFH) of the galaxy. A description of 
the SFH in terms of 15 age bins is too detailed given the effects of noise, 
intrinsic degeneracies of the synthesis process, and limited spectral coverage
of the data which is even further limited by the masks around the emission lines.
A coarser but more robust description of the SFH requires further binning of the age distribution.
Because of that we binned the results in 5 age bins: log(age)~$<$~8.3~dex, 8.3~$\leq$ log(age) $<$ 8.7 dex,
8.7 $\leq$ log(age) $<$ 9.3 dex, 9.3 $\leq$ log(age) $<$ 9.7 dex and log(age)~$\geq$~9.7~dex. This result is shown in Figure~\ref{starage}. 
The plot shows the flux fraction (x$_j$) - the fraction of light that
comes from the stellar population in that bin - as a function of age. The symbols represent
the fraction for each metallicity, and the black solid line represents the sum for
all three metallicities. 

From the synthesis
we obtained a light-weighted mean age of 0.8 Gyr. However, about 30\% of
the light comes from a population younger than 500 Myr. We also
obtained a light-weighted mean metallicity of 0.021 and 
Av = 0.80. The metallicities obtained for each stellar population
are somewhat puzzling, since it seems that the younger population is less metallic
than the older population. This trend however was observed before (\citealt{coelho09})
and it is probably an artifact from the method.

\subsection{Emission Line Measurements and Spectral Classification}

After the stellar population subtraction, the emission lines can be measured. 
We measured the emission lines
of NGC~185 after subtracting the stellar population in two ways: using NGC~205 and the synthetical spectra.
The first of these measurements were already done by HO97, but considering that
we are comparing results from different subtractions, we decided to do them again 
to avoid any bias from the method. In any case, the values we measured with 
the NGC~205 template subtraction agree very well with those obtained by HO97.
Results for the lines used in the classification are presented in Table 2. 
According to the classification
scheme presented in Table 1, the ratios obtained by the subtraction of NGC~205 place 
NGC~185 in the Seyfert category. However, a more careful analysis 
show that the [OIII]$\lambda$5007/H$\beta$ is in the borderline of the Seyfert classification
while [OI]$\lambda$6300/H$\alpha$ is actually in the LINER regime. This becomes clearer
when the stellar population subtraction is done with the stellar population synthesis.
Although visually there is not much difference from the spectra of NGC~205 and the synthetic
spectra, small differences can produce significant changes in the line ratios.
This happens because the absorption lines from the stellar population will affect 
more the hydrogen lines than the other lines. If these absorption lines are underestimated, 
the ratios used for the spectral classification will be higher than they should be. Using the synthetic spectra
to subtract the stellar population NGC~185 classification changes to LINER.

Since stellar population subtraction can drastically influence the 
line ratios, we wanted to determine an upper limit to the hydrogen absorption
line fluxes. As mentioned previously, younger stellar populations were detected in NGC~185. 
Our synthesis also confirms the presence of this population even though the fraction
obtained cannot be fully trusted.
The balmer absorption
lines are stronger in stellar populations of about 200 Myr. To test what is the maximum
effect of these lines in the emission lines measured, we also used a 200 Myr stellar population
as a template for the starlight of NGC~185, and subtracted it from the spectra. Results 
for this subtraction are also shown in Table 2. 
For this subtraction the classification becomes very uncertain. The ratio [OIII]$\lambda$5007/H$\beta$ is in the borderline of Seyfert classification, while the [OI]$\lambda$6300/H$\alpha$ ratio now points at a Transition Object.  The [NII]$\lambda$6300/H$\alpha$ and [SII]$\lambda$6300/H$\alpha$ ratios point to opposite classifications: The first is in the range of HII 
nuclei while the second would agree with any other active nuclei classification. 

In any of the three tests however, NGC~185 clearly has emission lines that cannot be explained 
simply by an HII region, which would not produce the line ratios observed. Although
the line ratios indicate some type of activity, it is very
unlikely that the emission lines are powered by an AGN, since it was not detected in many different surveys.
We cannot exclude the possibility of a heavily obscured AGN, although
the existence of a supermassive black hole should be much more 
common in massive bright galaxies. For galaxies with
absolute magnitude M$_B$ fainter than -18, this central massive
object would be most likely a compact stellar nucleus (Ferrarese et al. 2006). 
This is further supported by semi-analytical models which follow the formation and evolution
of black holes seeds formed at high redshift in the context of hierarchical cosmologies: 
low mass objects are more likely to eject their nuclear supermassive
black hole following a major merger as a result of gravitational recoil (Volontery et al. 
2007). To 
explain these line ratios we investigate other emission line mechanisms that could
be present in this galaxy.

\begin{table*}
\centering
%\begin{minipage}{80mm}
\caption{Emission-line intensity ratios used for spectral classification. }
\begin{tabular}{@{}cccccc@{}}
\hline
Subtraction   & H$\alpha$/H$\beta$ &[OIII]$\lambda$5007/H$\beta$ & [OI]$\lambda$6300/H$\alpha$ &[NII]$\lambda$6584/H$\alpha$  & [SII]$\lambda\lambda$6716,6731/H$\alpha$    \\
\hline
NGC 205            &  3.92  &  3.14   & 0.20 & 0.59 & 1.39    \\ 
Synthetic Spectra  &  4.72  &  2.53   & 0.25 & 0.54 & 1.31   \\
200 Myr population &  5.63  &  3.14   & 0.15 & 0.51 & 1.11 \\

\hline
\end{tabular}
%\end{minipage}
\label{ratios}
\end{table*}

%%%%%%%%%%%%%%%%%%%%%%%%%%%%%%%%%%%%%%%%%%%%%%%%%%%%%%%%%%%%%%%%%%%%%%%%%
\section{Can Planetary Nebulae fake a Seyfert Galaxy?}

NGC~185 has bright Planetary Nebulae (PNe) which were already identified in the Local
Group Census and previous studies (\citealt{corradi05} and references therein). Our group obtained deep spectroscopic
observations of the central region of NGC~185, aiming to 
study the H$\alpha$ emitting population (\citealt{goncalves11}).

GMOS-N pre-imaging exposures of a field of view of 5.5\arcmin $\times$ 5.5\arcmin\ at the central 
region of NGC~185 were taken in order to identify the PNe 
and other emission-line objects for the multi-object
spectroscopy. The two narrow-band frames were used to build a
\ha\ continuum-subtracted image, where we re-identified the 5 brightest PNe from \citet{richer95}, \citet{corradi05}, 
\citet{richer08}, together with other, much fainter, compact and 
diffuse emission-line objects. In total the selected 
objects for spectroscopy were 15, including the previously known PNe and new 
emission-line objects among them: PN, supernova remnant (SNR), 
and symbiotic system, and one diffuse object that is 
part of the arc-like central nebula, described as interestellar medium (ISM) emitting in H$\alpha$  by \citet{martinez99} and 
by \citet{corradi05}.
For details on the reduction process and the line measurements please refer to
\citet{goncalves11}.

\begin{figure} 
\includegraphics [width=85mm]{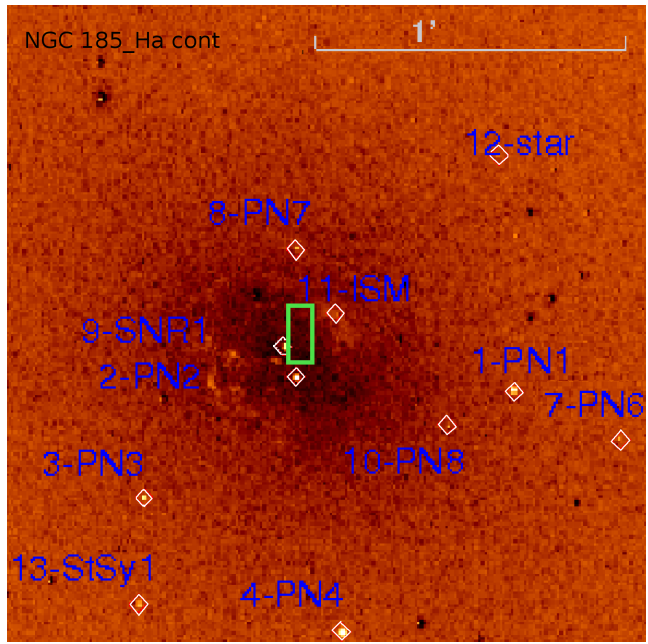}
\caption{  A zoom of the DSS2 image of NGC~185, with 2 $\times$ 2~arcmin$^2$,
and the location of the objects inside it that we studied spectroscopically. North is up 
and east points to the left. The original version of this figure with all the objects
we detected can be found in \citet{goncalves11}}
\label{ngc185image}
\end{figure}

Could the emission line of these objects be 
responsible for the emission lines observed in the integrated spectrum
of NGC~185 by HO97? 
Values of the line fluxes of interest are presented in Table 3, together with the fluxes 
measured by HO97. Four out of the 15 objects were stars and one is part of the diffuse 
nebula (ISM), so they
had no forbidden emission lines measured. We also do not present the fluxes for a Symbiotic star
we found, since it only has permitted lines observed. Off course, HO97 spectrum could not contain all these objects.
In fact, Figure~\ref{ngc185image} shows our objects in an image of NGC~185 and the 
slit of HO97 over it. 
Given the errors in the positions, caused by pointing inaccuracies 
(for example, GMOS pointing errors can be as high as 2"), it is probable 
that HO97 spectrum might have one or two objects only. However, this would be enough to explain at least the H$\beta$ flux they measured.  

\begin{table}
\centering
%\begin{minipage}{60mm}
{\scriptsize  
\caption{Observed fluxes normalized to $H\beta$ = 100 }
\begin{tabular}{@{}ccccccc@{}}
\hline
Object   & F$_{H\beta}$          &[OIII]         & [OI]            & H$\alpha$ &[NII] & [SII]    \\
          & $\times$ 10$^{-15}$  & $\lambda$5007 & $\lambda$6300   & &$\lambda$6584  &$\lambda\lambda$6716,6731 \\
  & erg cm $^-2$  &  &   &  & \\
\hline
PN1   & 1.46   &  653  &  3.34 & 368 &  41.8 &  4.15 \\ 
PN2   & 1.09   &  907  &  -    & 272 &   8.7 &  -    \\
PN3   & 0.72   & 1420  & 10.16 & 273 & 104.5 & 18.32 \\
PN4   & 3.33   & 1075  &  4.17 & 327 &  26.0 &  3.21 \\
PN5   & 0.19   & 1919  &  -    & 523 &  -    &  -    \\
PN6   & 0.05   & -     &  -    & 295 &  -    &  -    \\
PN7   & 0.35   & 2810  &  -    & 395 &  59.1 &  -    \\
SNR-1 & 0.04   & -     &  -    & 396 & 119.6 & 216.1 \\
PN8   & 0.07   & 1030  &  -    & 533 & 421   & 199.3 \\
%SySt-1& 0.22   &   -   &  -    & 431 &       &       \\
Total & 7.50   & 1058  &  3.5  & 335 & 38.0  & 6.9  \\
HO97 & 0.24 & 326 &  9.57 & 435 & 265   & 643     \\
\hline
\end{tabular}
}
%\end{minipage}
\label{fluxes}
\end{table}

The spectral classification is based on line ratios, and, although the PNe line
intensities could be high enough to justify the integrated emission observed in 
the galaxy, for most of them the line ratios would not explain the Seyfert classification,
and are very far from explaining the actual ratios measured by HO97. Table 4
shows the line ratios used for the classification for each object in our sample. 
Comparing these values with Table 1,
it is clear that the values for most objects are very far from what is expected for an active galaxy 
(Seyfert or LINER). For the PNe, while the [OIII] line is very strong, the low ionization lines
are too weak, specially [OI] and [SII]. Besides the PNe, we detected a resolved emission line
object close to the center of NGC~185, which has a {\it fwhm} two times
broader than the point spread function. This object (SNR-1 in Tables 3 and 4)
was identified in \citet{goncalves11}
as an old supernova remnant, and probably corresponds to the
object found by \citet{gallagher84}.
Its emission lines alone also cannot explain the Seyfert classification
of NGC~185, since it has no [OIII] emission. However, if we consider
in Figure~3 that both PN2 and SNR-1 contribute to HO97's slit, we might
have a different scenario. If we average these objects' line ratios,
weighted by their H$\beta$ fluxes, we find values of 
[OIII]$\lambda$5007/H$\beta$ = 9.89, [NII]$\lambda$6584/H$\alpha$ = 0.13
and [SII]$\lambda\lambda$6716,6731/H$\alpha$ = 0.04. These ratios are 
in agreement with the Seyfert regime. Although [OI] is still a problem here
and the values are very different from what was obtained by HO97, this gives
us a hint of what might be responsible for the emission lines observed in the 
integrated spectra of HO97.

\begin{table}
\centering
{\scriptsize  
%\begin{minipage}{80mm}
\caption{ Extinction corrected line ratios used in the spectral classification}
\begin{tabular}{@{}ccccc@{}}
\hline
 Object   &[OIII]$\lambda$5007& [OI]$\lambda$6300&[NII]$\lambda$6584  & [SII]$\lambda\lambda$6716,6731    \\
          &    /H$\beta$      & /H$\alpha$       &/H$\alpha$          &/H$\alpha$  \\
\hline
PN1   &  6.34  & 0.009 & 0.11 &  0.015 \\ 
PN2   &  9.07  &  -    & 0.03 &  -     \\
PN3   & 14.20  & 0.037 & 0.38 &  0.067 \\
PN4   & 10.57  & 0.013 & 0.08 &  0.010 \\
PN5   & 17.86  & -     & -    &  -     \\
PN6   &  -     & -     & 0.20 &  -     \\
PN7   & 26.25  & -     & -    &  -     \\
SNR-1 &  -     & -     & 0.33 & 0.533 \\
PN8   &  9.57  & -     & 0.78 & 0.358    \\
%SySt-1&  -     & -     & -    &  -\\

\hline
\end{tabular}
}
%\end{minipage}
\label{eclines}
\end{table}

%%%%%%%%%%%%%%%%%%%%%%%%%%%%%%%%%%%%%%%%%%%%%%%%%%%%%%%%%%%%%%%%%%%%%%%%%
\section{A Scenario with Supernova Remnants}

\cite{mnm10} took IRS (Infrared Spectrograph, on board of the Spitzer Space
Telescope) spectra of NGC 185, 
and that shows strong polycyclic aromatic hydrocarbon (PAH) emission, deep
silicate absorption features and H$_2$ pure rotational line ratios consistent with having the dust
and molecular gas inside the dust cloud being impinged by the far-ultraviolet radiation field
of a relatively young stellar population. This young stellar population was never really
detected, possibly due to the large extinction caused by the 
large amount of dust in this galaxy. Despite the fact that the current rate of star formation they 
measured is quite
low (∼ 10$^{-10}$ M$_\odot$/yr), this suggests that the star formation history of NGC 185 is complex.
 
Although it is very likely that a young stellar population is buried in this galaxy
and somehow contributing to the ionization process, it is very unlikely that this would
explain the high intensity of the low ionization emission lines. 

Like mentioned in the previous section, \citet{gallagher84}, 
took an integrated spectra of 
NGC~185, and they suggest that the emission lines could be explained by a supernova remnant.
They argue that the high intensities of [NII] and [SII] are characteristic of shock-heated
gas. The expected supernova (SN) rate in spheroidal galaxies is lower than in normal spiral galaxies. 
However, since this galaxy contains traces of young stars, the rate might be higher.

The spectral identification of supernova remnants (SNRs) was
pioneered in a series of papers by 
Mathewson and Clarke (1972, 1973a, 1973b, 1973c) where narrow-band optical interference
filters, centered on H$\alpha$ and the [SII]$\lambda$$\lambda$6717,6731 were
used to differentiate between primordial hydrogen and a heavy metal contaminated ejecta of a SNR. 
In this technique
the strength of [SII] lines should be about the same as the H$\alpha$ line, probably
due to shock fronts in the expanding SNR shell. The [SII] lines should
be at least an order of magnitude weaker than the H$\alpha$ line in HII regions as
compared to SNRs. Fesen, Blair \& Kirshner (1985) obtained optical spectra of galactic supernova remnants and
found that [OI]$\lambda$6300, [OII]$\lambda$3727 and [OIII]$\lambda$$\lambda$4959, 5007
are often all simultaneously strong in SNRs and this can be used to differentiate
SNRs from HII regions in cases where [SII]/H$\alpha$ is borderline.

\begin{figure*} 
\includegraphics [width=170mm]{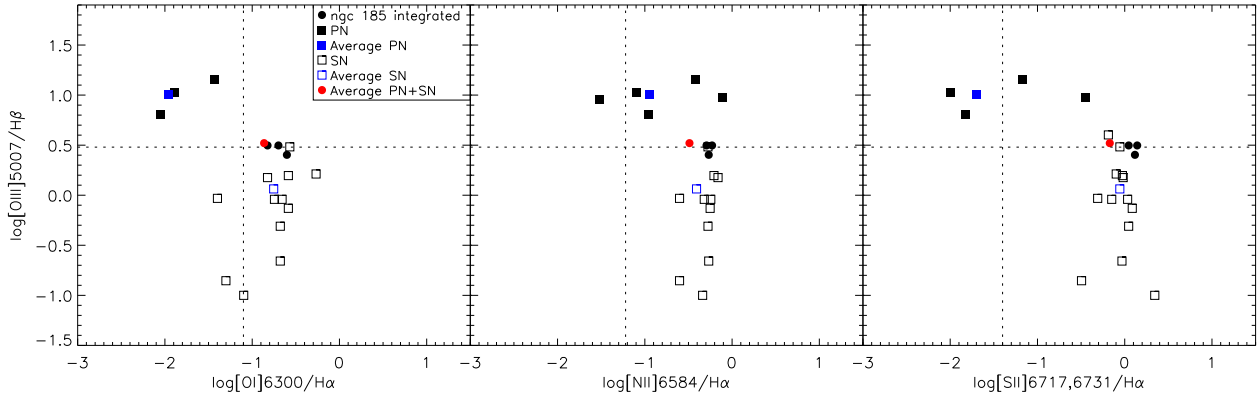}
\caption{ Diagnostic diagram for NGC~185. The spectral classification is based 
on the comparison of high excitation lines with low excitation ones. 
Dotted lines on these plots represent the division between active and non-active
galaxies, according to Table 1. Seyfert galaxies should be in the top right
corn of all these plots. }
\label{bptdiagram}
\end{figure*}

In our optical survey (\citealt{goncalves11}) we detected the presence of an old SNR close to the center of NGC~185, but if
more were present in the very central region, the detection in the optical would 
be very difficult due to the large extinction. However, the shock wave of this SN would
produce emission lines that could be detected in an integrated spectrum of the central region.
We can test if SNRs would be able to produce the line ratios
we see in the integrated spectrum of the galaxy. To do that we used Fesen et al. observations
of galactic SNRs, and compared their line ratios with what we measure for NGC~185. 
The results are shown in Figure~\ref{bptdiagram}. In this figure, the filled black circles are obtained from the
integrated spectra of HO97, and are the ratios we are trying to explain. Each of these points
represent a different stellar population subtraction. Filled black squares are the ratios corresponding 
to the PNe we observed, and the blue filled square is obtained averaging all their line fluxes and then 
calculating the ratios. For all three diagrams, it is clear that no combination of PNe could explain
the integrated line ratios. The open black squares represent line ratios for the galactic SNRs from 
Fesen et al. (1985). Their low ionization lines are much stronger than the ones from PNe, and
some of these objects come very close to the ratios we are trying to explain. 
One of them in particular explains all the lines in the three diagrams. If a
SNR with lines similar to this one would fall inside the slit, this would be enough
to explain NGC~185 lines. The blue open square in the diagrams
represents the average ratios for the SNRs.

If SNRs and PNe are present in NGC~185, it is possible, and even likely, that, in an integrated spectrum
of the central region,
both objects would fall inside the slit. What would be the result of this combination? To test that 
we took the average values for the PNe and for the SNRs, and calculated another average, weighted by
their median H$\beta$ flux (which are of the same order of magnitude). The result is the filled red
circle presented in Figure~\ref{bptdiagram}. This average has ratios very similar to the ones for the
integrated spectra of NGC~185. We then believe that the emission lines of NGC~185 are 
produced by a combination of PNe and SNRs, and that this combination fakes a Seyfert galaxy in a
BPT diagram.

%%%%%%%%%%%%%%%%%%%%%%%%%%%%%%%%%%%%%%%%%%%%%%%%%%%%%%%%%%%%%%%%%%%%%%%%%
\section{Predicted SN Rates from Chemical Evolution Models}

Although young SNRs were not detected in the nuclear region of NGC~185, we can 
investigate if they are expected to be present in this galaxy. 
Since the star formation history inferred in this work and others 
(Mateo 1998; \citealt{martinez99}; Dolphin et al. 2005) indicates a very recent star formation episode in NGC~185, it is reasonable to think that SNe should have exploded recently in the galaxy. The rate at which SNe occur, however, is crucial to determine whether or not they could be responsible for the emission lines observed. A very low rate would probably mean that the contribution of the SNRs for the emission lines is negligible whereas a high rate would be a strong indication that SNRs are indeed present in NGC~185, even if they are not observed.

The SNe rate of a galaxy can be predicted by detailed chemical evolution models. 
After a maximum mass, a star formation history (SFH), an initial mass function (IMF), and a star formation rate are chosen as inputs of the model, the code solves the basic equations of chemical evolution (Tinsley 1980, Matteucci 1996) and calculates the evolution in time of the mass fraction of several chemical species in the gas of the galaxy by taking into account the contribution of stars from 1 to 100 solar masses and their lifetimes. The contribution of the stars to the abundance of the chemical elements (stellar yields) and to the energy (in the form of stellar winds and supernovae type Ia and type II) of the interstellar medium are considered. The stellar yields play an important role in the abundances of chemical elements whereas the stellar winds and the SNe explosion, together with the dark matter content, define the occurrence of galactic winds (Bradamante, Matteucci, D'Ercole, 1998). Normally, the star formation history adopted in the models are the ones derived from observations and the star formation rate, IMF, and efficiency of galactic winds are constrained strongly by the abundance ratios, the stellar metallicity distribution, and the age-metallicity relation observed. Once the main parameters of the model are adjusted to fit the data, the SNe rate comes naturally, given the stellar lifetimes, as an output of the code.

\begin{figure} 
\includegraphics [scale=0.5]{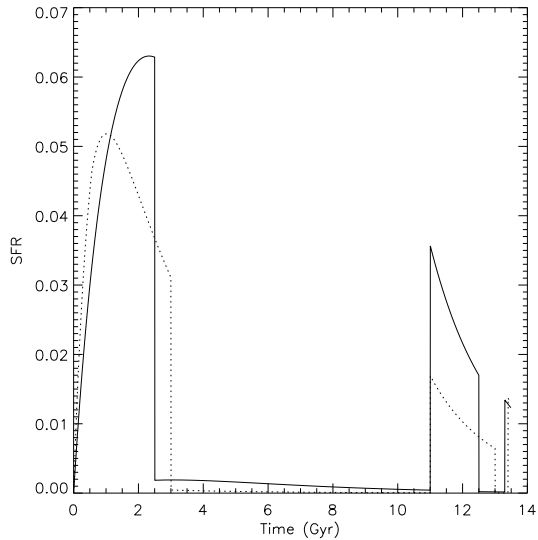}
\caption{The SFHs adopted by the chemical evolution models for NGC~185. The dotted line represents model 1 and the solid line represents model 2.}
\label{sfr}
\end{figure} 

In this work, we adopted the chemical evolution code of Lanfranchi $\&$ Matteucci (2003, 2004), developed to fit several observed data of dwarf irregular and dwarf spheroidal galaxies. The most important parameters in the models are the SFH, the SF efficiency and the galactic wind efficiency. Dwarf Irregular galaxies are normally characterized by 
several short bursts of star formation with high rates and low galactic wind efficiencies (Yin et al. 2010, Yin, Matteucci \& Vladilo 2011) whereas the dwarf spheroidals exhibit a few long (few Gyr) episodes of SF with low rates and very efficient winds (Lanfranchi $\&$ Matteucci 2004). The star formation histories adopted for the NCG~185 models are generally the same as the one inferred in this work (Figure~\ref{starage}): three episodes of activity separated by long intervals. The first episode is the most intense and starts at the beginning of the formation of the galaxy lasting a few Gyr (from 3 to 4), the second one occurs between 10 to 13 Gyr with a lower rate compared to the first, and the last one, in the last few hundred Myr, has the lowest efficiency (Figure~\ref{sfr}). 
The two curves in the figure represent two models with slightly different SFH
(solid line - model 1, dotted line - model 2). The others parameters (IMF, total mass, infall timescale) are essentially the same in both models and identical to the ones adopted in the dwarf models from Lanfranchi $\&$ Matteucci (2003, 2004).
The predictions of the models were compared to the chemical properties Of NGC~185, specially the abundance ratios (determined for three PNe), the age-metallicity relation, and the total present day mass. The IMF follows the prescriptions 
suggested by Salpeter (1955), the stellar yields are the same as in Lanfranchi \& Matteucci (2010), and the efficiency of the galactic wind is low ($w_i$ = 0.4 - 0.6). For more details of the code we refer the reader to Lanfranchi \& Matteucci (2003, 2010).

\begin{figure} 
\includegraphics [scale=0.5]{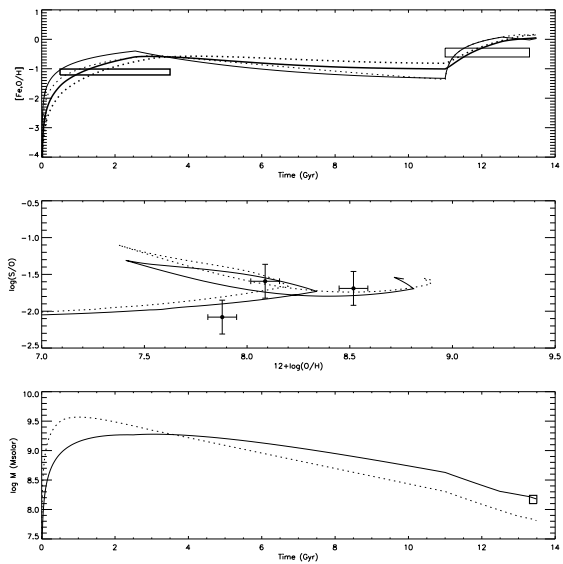}
\caption{Comparison between the predictions of the chemical evolution model for NGC~185 and observed data. Upper panel - [Fe,O/H] vs. time, middle panel - log(S/O) vs (O/H), lower panel - total mass vs. time. The lines are the same as in Figure~\ref{sfr}. The large squares in the upper and lower panel and the dots in the middle one represent observed values.}
\label{data}
\end{figure}

The exact time of occurrence of the SF episodes and their duration is not certain, as mentioned before. Because of that, we varied slightly the SFH given in section 2, trying to find the best fit to the observed data. The two best models are models 1 and 2 in Figure ~\ref{sfr}. The most important feature of both SFHs is the occurrence of three episodes of activity 
with a long quiescent interval (a few Gyrs) between the first two episodes and a 
shorter interval between the second and the last episode. During the first SF 
episode the metallicity of the galaxy increases fast reaching [Fe/H] = -0.6 dex in $\sim 2.5$~Gyr and then decreases slowly in the inter-burst period due to the inactivity and to 
the galactic wind. The metallicity starts increasing again with the second and third episodes of SF, reaching [O/H] = -0.6 dex at an age of $\sim$ 12 Gyr. The age-metallicity relation predicted by both models is in agreement with the data, as one can notice in Figure~\ref{data} (upper panel). The lines are the same as in Figure~\ref{sfr} and the large squares represent the observed values of [Fe/H] in stars (thick line - Butler $\&$ Martinez-Delgado 2005) and [O/H] in PNe (thin line - \cite{goncalves11}). In order to make a proper comparison with the data it is shown the predictions of the models for [Fe/H] (thick line) and [O/H] (thin line).

Besides the age-metallicity relation, the predictions of the models are compared to the 
observed (S/O) ratios. Nitrogen and helium were not considered because their abundances in PNe may not represent the abundance of the interstellar medium at the epoch of the formation of the star responsible for the PNe (see Chiappini et al. 2009). The evolutionary tracks of the models reproduce very well the (S/O) ratios observed in all three PNe (Figure~\ref{data} - middle panel). The loop in the track is caused by the galactic wind that reduces considerably (O/H) increasing (S/O) at the same time. Small changes in the SFH and SF efficiency do not change considerably the results as one can see by th lines of the two different models in the Figure~\ref{data}. The galactic wind plays an important role also in the final luminous mass of the galaxy. A galactic wind with a low rate of gas removal does not change much the mass of the galaxy, since only a small fraction of gas is lost. Our choice of wind efficiency allows us to reproduce the age-metallicity relation, the (S/O) ratio, and also the present day total mass of the galaxy (Figure~\ref{data} - lower panel). Battinelli $\&$ Demmers (2004) inferred a total mas of 1.3 x $10^8 M_{\odot}$ (small square in the lower panel of Figure~\ref{data}) whereas the models predict values between 0.7 - 1.5 x $10^8 M_{\odot}$ (models 1 and 2, respectively).

\begin{figure} 
\includegraphics [scale=0.5]{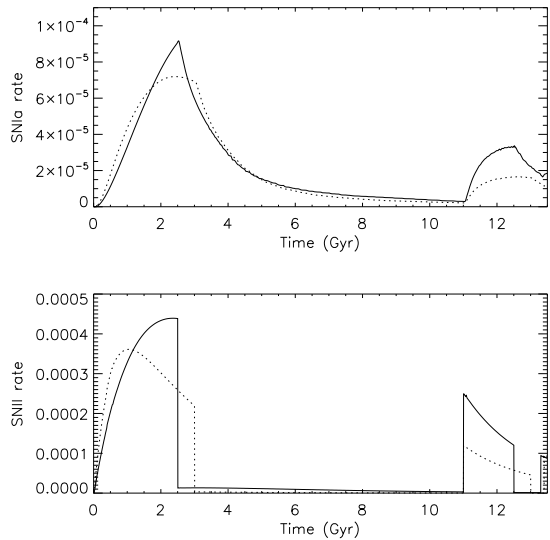}
\caption{The SNe rate predicted by the chemical evolution model for NGC~185. The lines are the same as in Figure~\ref{sfr}.}
\label{snerate}
\end{figure}

After adjusting the main parameters of the models to reproduce the observational constraints, it is possible to infer the SNe (types Ia and II) rate and the number of SNe that might have exploded recently. The predictions of models 1 and 2 for SNII (up panel) and SNIa (lower panel) rates in number of explosions per century are shown in Figure~\ref{snerate}. Even though the absolute number is low in both cases (0.88-0.94 x $10^{-4}$ century $^{-1}$ for SNII and 0.11-0.18 x $10^{-4}$ century$^{-1}$ for SNIa) one should consider that SNII could have exploded during all the period of the last SF episode, being distributed in time according to the stellar lifetimes, and that SNIa occurred over an even longer period due to the typical lifetimes of their progenitors. In that sense, we can estimate the number of SNe explosion of both types by considering the time intervals within which the rates of explosions are not negligible. The SNe II rate is very low in the period between $\sim$ 1.5 to 0.4 Gyr ago until a SF episode took place. With the formation of new stars, new SNII starts exploding again increasing the SNe II rate, that remains high until now. The rate of SNIa, on the other hand, although lower, remains considerable for the last $\sim$ 3 Gyr. Taking that into consideration, we can estimate that around 80 SNe II and 15 SNIa exploded in the last 100 Myr.

%%%%%%%%%%%%%%%%%%%%%%%%%%%%%%%%%%%%%%%%%%%%%%%%%%%%%%%%%%%%%%%%%%%%%%%%%
\section{Summary and conclusions}

In this paper we investigate the spectral classification of the dwarf
spheroidal galaxy NGC~185 as a Seyfert galaxy. This classification 
was defined by HO97 based on the integrated optical spectrum of the
central 2"x4" of the galaxy. The emission line ratios measured by
them places this galaxy on the Seyfert category.
However, NGC~185 was not detected in radio surveys either in 6 cm
or 20 cm, or x-ray observations. Although the possibility
of a heavily obscured AGN cannot be excluded by
our data, the existence of a supermassive black hole in the center
of a small, low-luminosity galaxy as NGC~185 is very unlikely.
In the absence of an active nucleus, the Seyfert-like
line ratios have to be produced by some other process, probably stellar.

Since the emission line measurements are very dependent on
the stellar population subtraction, we performed stellar population
synthesis on the integrated spectrum from HO97 to test how much
the emission line ratios measured were dependent on the methods
used. Although the ratios change depending on the method used for
the stellar population subtraction, these changes are too small to justify
the misclassification. This means that the emission lines measured are
indeed there and are not an artifact of the method.

It is known that NGC~185 has bright PNe, and our group recently
performed a deep spectroscopy observation of this galaxy searching
for the H$\alpha$ emitting population. Although the fluxes of individual
objects are high enough to explain the emission lines observed
in the integrated spectra, the line ratios of the PNe are very far
from what was measured for NGC~185. [OIII]$\lambda$$\lambda$4959, 5007/ H$\alpha$
is too high and the low ionization lines are too low for these objects. 
The emission lines of these objects might be contributing to the integrated spectrum,
but something else is needed to explain the low ionization lines. 

Besides the PNe, we found two other H$\alpha$ emitters: a SNR and a symbiotic star.
SNRs are known to have very strong low ionization lines like [OI], [NII] and
[SII] lines. The one we detected is an old object, since its [SII] and [NII] are low
compared with average SNRs, and no [OI] was detected. If we consider the two objects
closer to the center (SNR-1 and PN2) and calculate their average line ratios, 
the results would place NGC~185 in the Seyfert category. The [OI] is however, the
hardest to explain.

Young SNRs usually have strong [OI] emission. We then consider the possibility
that there is a young SNR in the central part of NGC~185, which was not detected 
due to the extinction caused by the large amount of dust in this
region. It was already suggested that a young stellar population 
is present in the central part of the galaxy, which was confirmed by our
stellar population synthesis: our results suggest that 30\% of the light
might be coming from a population younger than 500 Myr. 
The same young population originate a considerable number of SNe explosions in the last 100 Myr, 
as predicted by chemical evolution models that adopted the SFH derived 
by the stellar population synthesis. These models
reproduce the observed (S/O) ratios, the total present-day mass, and the age-metallicity relation. The models pointed that $\sim$ 90 SNe exploded recently, indicating a large probability that at least a few young SNR should be present in the galaxy. 
To demonstrate how a young SNR could explain the line ratios observed in the galaxy we
used a sample of galactic SNRs from \cite{fea85}. The combination 
of the average line fluxes
of these objects with the average line fluxes of the PNe we found gives
line ratios that are exactly in the range of the ratios measured for 
the integrated spectrum of NGC~185. We then conclude that the emission
lines observed in the integrated spectrum of HO97 are probably result of 
either one SNR or
the combination of a PN and a SNR inside their slit. 
As stated previously, we cannot exclude the possibility of an
obscured AGN, but we believe it is very unlikely that NGC~185 is a Seyfert galaxy.

%%%%%%%%%%%%%%%%%%%%%%%%%%%%%%%%%%%%%%%%%%%%%%%%%%%%%%%%%%%%%%%%%%%%%%%%%
\section*{Acknowledgments} 
L.P.M. thanks the financial support by the Brazilian agency FAPESP (2011/00171-4). 
L.P.M. also thanks L'Oreal Brasil and ABC for finnantial support. D.R.G.
kindly acknowledges the UCL Astrophysics Group for their hospitality. LM is
supported through the ASIINAF grant ``HeViCS: The Herschel Virgo Cluster
Survey'' I/009/10/0. The work of CQ is supported by the INCT-A (PDJ 154908/2010-0).


\begin{thebibliography}{99}

\bibitem[\protect\citeauthoryear{Asari et al.}{2007}]{asari07} 
Asari, N.V., Cid Fernandes, R., Stasi\'nska, G., Torres-Papaqui, J.P., Matheus, A., Sodr\'e, L., Schoenell, W., Gomes, J.M. 2007, MNRAS, 381, 263

\bibitem[\protect\citeauthoryear{Baade}{1951}]{baade51}
Baade W., 1951, Publ. Obs. Univ. Michigan, 10, 7

\bibitem[\protect\citeauthoryear{Baldwin, Phillips \& Terlevich}{1981}]{bpt81}
Baldwin, J.A., Phillips, M.M., Terlevich, R. 1981, PASP, 93, 5

\bibitem[\protect\citeauthoryear{}{}]{bp04} Battinelli, P., Demmers, S., 2004, A\&A, 417, 479

\bibitem[\protect\citeauthoryear{}{}]{bmd98}Bradamante F., Matteucci F., D'Ercole A., 1998, A\&A, 337, 338

\bibitem[\protect\citeauthoryear{Brandt et al.}{1997}]{brandt97}
Brandt, W.N., Ward, M.J., Fabian, A.C., Hodge, P.W. 1997, MNRAS, 291, 709

\bibitem[\protect\citeauthoryear{Bruzual \& Charlot}{2003}]{bc03}
Bruzual, G., Charlot, S. 2003, MNRAS, 344, 1000

\bibitem[\protect\citeauthoryear{}{}]{bm05} Butler, D. J., Mart\'inez-Delgado, D., 2005, AJ, 1129, 2217

\bibitem[\protect\citeauthoryear{}{}]{cgsb} Chiappini, C., G\'orny, S. K., Stasi\'nska, G., Barbuy, B., 2009, A\&A, 494, 591

\bibitem[\protect\citeauthoryear{Cid Fernandes et al.}{2004}]{cea04} 
Cid Fernandes, R., Gu, Q., Melnick, J., Terlevich, E., Terlevich, R., Kunth, D., Rodrigues Lacerda, R., Joguet, B. 2004, MNRAS, 355, 273

\bibitem[\protect\citeauthoryear{Cid Fernandes et al.}{2005}]{cea05}
Cid Fernandes, R., Matheus, A., Sodr\'e, L., Stasi\'ska, G., Gomes, J.M. 2005, MNRAS, 358, 363

\bibitem[\protect\citeauthoryear{Cid Fernandes et al.}{2009}]{cea09}
Cid Fernandes, R. , Schoenell, W., Gomes, J.M., Asari, N.V., Schlickmann, M., Mateus, A., Stasinska, G., Sodré, L.,Jr., Torres-Papaqui, J.P. 2009, RMxAC, 35, 127

\bibitem[\protect\citeauthoryear{Coelho, Mendes de Oliveira \& Cid Fernandes}{2009}]{coelho09}
Coelho, P., Mendes de Oliveira, C., Cid Fernandes, R. 2009, MNRAS, 396, 624

\bibitem[\protect\citeauthoryear{Corradi et. al.}{2005}]{corradi05}
Corradi R. L. M., Magrini L., Greimel R., Irwin M., Leisy P., et al., 2005, A\&A, 431, 555

\bibitem[\protect\citeauthoryear{}{}]{dwsh} Dolphin, A.E., Weisz, D.R., Skillman, E.D., Holtzman, J.A., 2005,
to appear in Valls-Gabuad D. \& Chavez M., eds.,
Resolved Stellar Populations, ASP Conference Series, astro-ph/0506430

\bibitem[\protect\citeauthoryear{}{}]{f06} Ferrarese, L., C\^oté, P., Dalla Bont\`a, E., 
Peng, E.W., Merritt, D., Jord\'an, A., Blakeslee, J.P., Hasegan, M., Mei, S., 
Piatek, S., Tonry, J.L.m West, M.J. 2006, ApJL, 644, L21

\bibitem[\protect\citeauthoryear{Fesen et al.}{1985}]{fea85}
Fesen, R.A., Blair, W.P., Kirshner, R.P. 1985, ApJ, 292, 29 

\bibitem[\protect\citeauthoryear{Gallagher et al.}{1984}]{gallagher84}
 Gallagher  J. S. III, Hunter D. A., \& Mould J. R., 1984, ApJ, 281, L63

\bibitem[\protect\citeauthoryear{Gon\c calves et al.}{2011}]{goncalves11}
Gon\c calves D. R., Magrini L., Martins, L.P., Teodorescu, A.M., Quireza, C. 2011,
MNRAS, in press (astroph-1109.3019)

\bibitem[\protect\citeauthoryear{Heckman, Crane \& Balick}{1980}]{hbc80}
Heckman, T., Crane, P.C., Balick, B. 1980, A\&AS, 40 295

\bibitem[\protect\citeauthoryear{Ho et al.}{1997}]{ho97}
Ho, L., Filippenko, A.V., Sargent, W.L.W. 1997, ApJ, 112, 315

\bibitem[\protect\citeauthoryear{Ho \& Ulvestad}{2001}]{houlv01}
Ho, L., Ulvestad, J.S. 2001, ApJS, 133, 77

\bibitem[\protect\citeauthoryear{}{}]{lm03} Lanfranchi G., Matteucci F., 2003, MNRAS, 345, 71

\bibitem[\protect\citeauthoryear{}{}]{lm04} Lanfranchi, G., \& Matteucci, F., 2004, MNRAS, 351, 1338

\bibitem[\protect\citeauthoryear{}{}]{lm10} Lanfranchi, G., \& Matteucci, F., 2010, A\&A, 512A, 85

\bibitem[\protect\citeauthoryear{Lee et al.}{1993}]{lea93}
Lee, M.G., Freedman, W.L., Madores, B.F. 1993, AJ, 106, 964

\bibitem[\protect\citeauthoryear{Marleau et al.}{2010}]{mnm10}
Marleau, F.R., Noriega-Crespo, A., Misselt, K. 2010, ApJ, 713, 992

\bibitem[\protect\citeauthoryear{Mart\'\i nez-Delgado et al.}{1999}]{martinez99}
Mart\'i nez-Delgado D., Aparicio A., \& Gallart C., 1999, AJ, 118, 2229

\bibitem[\protect\citeauthoryear{}{}]{m96} Matteucci F., 1996, FCPh, 17, 283

\bibitem[\protect\citeauthoryear{Mateo}{1998}]{mateo98}
Mateo M. L., 1998, ARA\&A 36, 435

\bibitem[\protect\citeauthoryear{Mateus et al. }{2006}]{mateus06}
Mateus, A., Sodr\'e, L, Cid Fernandes, R., Stasi\'nska, G., Schoenell, W., Gomes, J.M. 2006, MNRAS, 370, 721

\bibitem[\protect\citeauthoryear{Mathewson \& Clarke}{1972}]{mc72}
Mathewson, D.S., Clarke, J.N. 1972, ApJL, 178, 105

\bibitem[\protect\citeauthoryear{Mathewson \& Clarke}{1973a}]{mc73a}
Mathewson, D.S., Clarke, J.N. 1973a, ApJL, 179, 89

\bibitem[\protect\citeauthoryear{Mathewson \& Clarke}{1973b}]{mc73b}
Mathewson, D.S., Clarke, J.N. 1973b, ApJL, 180, 725

\bibitem[\protect\citeauthoryear{Mathewson \& Clarke}{1973c}]{mc73c}
Mathewson, D.S., Clarke, J.N. 1973c, ApJL, 182, 697

\bibitem[\protect\citeauthoryear{Richer \& McCall}{1995}]{richer95}
Richer M. G., \& McCall M. L.,  1995, ApJ, 445, 642

\bibitem[\protect\citeauthoryear{Richer \& McCall}{2008}]{richer08}
Richer M. G., \& McCall M. L.,  2008, ApJ, 684, 1190

\bibitem[\protect\citeauthoryear{}{}]{s55} Salpeter E.E., 1955, ApJ, 121, 161

\bibitem[\protect\citeauthoryear{}{}]{t80} Tinsley B.M., 1980, FCPh, 5, 287

\bibitem[\protect\citeauthoryear{Veilleux \& Osterbrock}{1987}]{vo87} 
Veilleux, S., Osterbrock, D.E. 1987, ApJS, 63, 295

\bibitem[\protect\citeauthoryear{}{}]{v07}	Volonteri, M., Sikora, M., Lasota, J.-P.
2007, ApJ, 667, 704

\bibitem[\protect\citeauthoryear{}{}]{ymmlg} Yin, J., Magrini, L., Matteucci, F., Lanfranchi, G. A., Gonçalves, D. R., Costa, R. D. D., 2010, A\&A, 520A, 55

\bibitem[\protect\citeauthoryear{}{}]{ymv} Yin, J.,  Matteucci, F.,  Vladilo, G., 2011, A\&A, 531A, 136

\end{thebibliography}
\end{document}